\documentclass{aastex}
\usepackage{spr-astr-addons}
\usepackage{url}

\RequirePackage{color}

\newcommand{\emaila}{gtrcosmo@gmail.com}

\begin{document}

\title{LRS Bianchi I model with constant expansion rate in $f(R,T)$ gravity}
\shorttitle{LRS Bianchi I model with constant expansion rate in $f(R,T)$ gravity}
\shortauthors{V. Singh and A. Beesham}

\author{Vijay Singh\altaffilmark{1}} \email{\emaila} \and \author{Aroonkumar Beesham\altaffilmark{2}}
\affil{Department of Mathematical Sciences,\\
 University of Zulualnd, Private Bag X1001,\\
 KwaDlangezwa, South Africa - 3886}

\altaffiltext{1}{gtrcosmo@gmail.com}
\altaffiltext{2}{abeesham@yahoo.com}

\begin{abstract}
An LRS Bianchi-I space-time model is studied with constant Hubble parameter in $f(R,T)=R+2\lambda T$ gravity. Although a single (primary) matter source is considered, an additional matter appears due to the coupling between matter and $f(R,T)$ gravity. The constraints are obtained for a realistic cosmological scenario, i.e., one obeying the null and weak energy conditions. The solutions are also extended to the case of a scalar field (normal or phantom) model, and it is found that the model is consistent with a phantom scalar field only. The coupled matter also acts as phantom matter. The study shows that if one expects an accelerating universe from an anisotropic model, then the solutions become physically relevant only at late times when the universe enters into an accelerated phase. Placing some observational bounds on the present equation of state of dark energy, $\omega_0$, the behavior of $\omega(z)$ is depicted, which shows that the phantom field has started dominating very recently, somewhere between $0.2\lesssim z\lesssim0.5$.
\end{abstract}

\section{Introduction}\label{s:intro}

\cite{Harkoetal2011}  proposed a general non-minimal coupling between matter and geometry in the framework of an effective gravitational Lagrangian consisting of an arbitrary function of the Ricci scalar $R$, and the trace $T$ of the energy-momentum tensor, and introduced $f(R,T)$ gravitational theory. An extra acceleration in $f(R,T)$ gravity results not only from a geometrical contribution, but also from the matter content. This extraordinary phenomena of $f(R,T)$ gravity may provide some significance signatures and effects which could distinguish and discriminate between various gravitational models. Therefore, this theory has attracted many researchers to explore different aspects of cosmology and astrophysics in isotropic and as well as in anisotropic space-times (see for example \cite{Jamiletal2012,Reddyetal2013,Azizi2013,Alvarengaetal2013jmp4,Alvarengaetal2013prd87,Sharifetal2013epjp128,
Chakraborty2013,Houndjoetal2013cjp91,Pasquaetal2013,RamPriyankaASS2013,SinghSinghASS2015,Baffouetal2015,
BaffouetalPRD2018,SantosFerstMPLA2015,NoureenetalEPJC2015,ShamirEPJC2015,SinghSingh2016,
AlhamzawiAlhamzawiIJMPD2016,Yousafetal2016,Alvesetal2016,ZubairetalEPJC2016,SofuogluASS2016,
MomenietalASS2016,DasetalEPJC2016,SalehiAftabiJHEP2016,SahooetalEPJC2018,MoraesetalEPJC2018,SinghBeeshamEPJC2018,
SrivastavaSinghASS2018,SharifAnwarASS2018,TiwariBeeshamASS2018,ShabaniZiaieEPJC2018,RajabiNozariPRD2017,
MoraesetalIJMPD2019,DebetalMNRAS2019,LobatoetalEPJP2019,BaffouetalPRD2018,DebetalMNRAS2019,TretyakovEPJC2018,
ElizaldeKhurshudyanPRD2018,OrdinesCarlsonPRD2019,MauryaTello-OrtizbJCAP2019,DebetalMNRAS2019,
EsmaeiliJHEP2018} and references therein).

The first work on any anisotropic model in $f(R,T)$ gravity was done by \cite{AdhavASS2012} in LRS Bianchi I space-time. The author considered a particular form of $f(R,T)=R+2\lambda T$, where $\lambda$ is an arbitrary constant, and obtained the solutions by assuming a constant expansion rate. This assumption corresponds to the accelerating expansion of the universe. A serious shortcoming in his work is, due to an incorrect field equation, the solutions are mathematically and physically invalid. Our purpose in this paper is to address the correct field equations and explore the geometrical and physical properties of this model.

If one considers any matter in this theory, then due to the coupling between matter and $f(R,T)$ gravity, some extra terms appear on the right hand side of the field equations. These terms must be treated as matter as well and may be called  coupled matter. It may act either as a perfect fluid or DE. Therefore, the effective matter in these models is a sum of primary matter and coupled matter. One may ensure a physically viable scenario by demanding the weak energy condition (WEC)\footnote{$\rho\geq0$, $\rho+p\geq0$} for the primary matter and coupled matter.  We have followed this criteria in our recent study \cite{SinghBeeshamEPJP2020}. We shall follow this criteria in present work too.

In addition to that we also study the physical viability of the model through the energy conditions, and find the constraints for a realistic cosmological scenario. Further, we extend our solutions to the case of a normal/phantom scalar field model to determine the nature of the matter. We also study the behavior of the matter by using the present values of the equation of state parameter consistent with observational constraints. We also examine the role of $f(R,T)$ gravity in this study.

The work is organised as follows. In Sec. 2 we show that the geometrical behavior of the model reported by \cite{AdhavASS2012} is independent of $f(R,T)$ gravity. In Sec. 3 we present the correct field equations for an LRS Bianchi I spacetime model in $f(R,T)=R+2f(T)$ gravity, and find the constraints for a realistic physical scenario ensuring positivity of the energy density. The scalar field model is considered in Sec. 3.1 followed by a study of the behavior of the effective matter through the equation of state parameter. The findings are summarised in Sec. 4. Note that the equation numbers in  round brackets throughout our discussion refer to the equations of our work, whereas the equation numbers in round brackets including the section numbers and some points number mentioned within inverted commas refer to the Ref. \cite{AdhavASS2012}.

Although we consider a single matter source in our model, an extra matter source appears in $f(R,T)=R+2\lambda T$ gravity due to the coupling terms of $T$ with the  matter on the right hand side of the field equations. First, we obtain the constraints for the primary matter to obey the weak energy condition. This not only ensures a realistic cosmological scenario, but also helps to identify the various evolutionary phases of the universe, specifically, it distinguishes between early inflation and late time acceleration. Anticipating the dual nature of primary matter, we replace it with a scalar field to know the actual nature (perfect fluid, quintessence or phantom) of primary matter. Further, we analyze the behavior of coupled matter in the various evolutionary phases. In this way,  we determine which matter source causes inflation, deceleration and late time acceleration, and what is the role of $f(R,T)$ gravity in the course of evolution of the universe.

\section{The solutions in general relativity}

In this section we show that the geometrical behaviour in points ``(i)--(iv)" addressed by \cite{AdhavASS2012} in section ``4", is independent from  $f(R,T)$ gravity, and it remains similar to that in general relativity (GR).

A spatially homogenous and anisotropic locally-rotationally-symmetric (LRS) Bianchi I space-time metric is given by
\begin{equation}
ds^{2} =dt^{2}-A^2(t)dx^2-B^2(t)(dy^2+dz^2),
\end{equation}
where $A$ and $B$ are the scale factors, and are functions of cosmic time $t$.

The average scale factor for the metric (1) is defined as
\begin{equation}
  a=(AB^2)^{\frac{1}{3}}.
\end{equation}
The average Hubble parameter (average expansion rate) $H$, which is the generalization of the Hubble parameter in the isotropic case, is given by
\begin{equation}
  H=\frac{1}{3}\left(\frac{\dot A}{A}+2\frac{\dot B}{B}\right).
\end{equation}
Consider the energy-momentum tensor
\begin{equation}
  T_{\mu\nu}=(\rho+p)u_\mu u_\nu-pg_{\mu\nu},
\end{equation}
\noindent where $\rho$ is the energy density and $p$ is the thermodynamical pressure of the matter. In comoving coordinates $u^\mu=\delta_0^\mu$, where $u_\mu$ is the four-velocity of the fluid which satisfies the condition $u_\mu u^\nu=1$.

The Einstein field equations read as
\begin{equation}
R_{\mu\nu}-\frac{1}{2}R g_{\mu\nu}=T_{\mu\nu},
\end{equation}
where the system of units $8\pi G=1=c$ are used.

The above field equations for the metric (1) and energy-momentum tensor (4), yield
\begin{eqnarray}
  \left(\frac{\dot B}{B}\right)^2+2\frac{\dot A\dot B}{A B}&=&\rho,\\
\left(\frac{\dot B}{B}\right)^2+2\frac{\ddot B}{ B}&=&-p,\\
\frac{\ddot A}{A}+\frac{\ddot B}{B}+\frac{\dot A \dot B}{AB}&=&-p.
\end{eqnarray}
where a dot denotes the derivative with respect to cosmic time $t$. These are three independent equations with four unknowns, namely $A$, $B$, $\rho$ and $p$. Therefore, one requires a supplementary constraint to find the exact solutions of the field equations. \cite{AdhavASS2012} considered the case of a constant expansion rate
\begin{equation}
H=k,
\end{equation}
where $k>0$ is a constant. Since $H=\dot a/a$, the average scale factor evolves as
\begin{equation}
  a(t)=a_0e^{k t},
\end{equation}
where $a_0$ is an integration constant.

The deceleration parameter, $q=-a\ddot a/\dot a^2=-1-\dot H/ H^2$ takes a constant value
\begin{equation}
q=-1,
\end{equation}
which corresponds to an accelerating expansion of the universe.

From (7) and (8), one has
\begin{equation}
\frac{\dot A}{A} -\frac{\dot B}{B}=\frac{\beta}{AB^2},
\end{equation}
where $\beta$ is a constant of integration.

From (3) and (12), by the use of (9), one obtains
\begin{eqnarray}
   A&=&c_1e^{kt-\frac{2\beta e^{-3kt}}{9k}},\\
B&=&c_1e^{kt+\frac{\beta e^{-3kt}}{9k}},
\end{eqnarray}
where $c_1$ is a constant of integration and another integration constant is taken as  unity without any loss of generality. In section ``3", namely, ``physical properties", \cite{AdhavASS2012} worked out some kinematical parameters and also obtained the expressions for the energy density and pressure. The author in his conclusion mentioned that the scale factors are the solutions of the LRS Bianchi I model in $f(R,T)$ gravity. But here one can see that the scale factors (13) and (14) are obtained in GR. Hence, the behaviour of the kinematical parameters, namely, the expansion scalar, shear scalar, and the anisotropy parameter discussed by Adhav remain is independent of $f(R,T)$ gravity and remain the same as in GR. In our recent work (\cite{SinghBeeshamGRG2019}), we have explored the features of these parameters.

The main issue in Adhav's paper is, the expressions for the energy density and pressure are incorrect due to a wrong field equation, namely, equation number ``(2.5)". Therefore, the solutions obtained by the author are invalid mathematically. In the next section, we shall reformulate this model. We find the constraints for a physically realistic cosmological scenario and explore the physical behavior of the model. We shall also extend the solutions to a scalar field model.

\section{The solutions in $f(R,T)$ gravity}

In Sect. 2, $\rho$ and $p$, respectively, are the effective energy density and pressure in the model of GR.  When one we considers the energy-momentum tensor (4) in $f(R,T)$ gravity then $\rho$ and $p$ no longer correspond to the effective energy density and pressure. As aforementioned in the introduction that due to the coupling between matter and trace some extra terms appear in the right hand side of the field equation in $f(R,T)$ gravity. These terms can also be treated as matter. We may call it coupled matter. The matter given by the energy-momentum tensor (4) should be treated as the primary matter. Let us replace $\rho$ and $p$ with $\rho_m$ and $p_m$, respectively, which represent the energy density and pressure of primary matter.  The notations for coupled matter are defined in Sect. 3.2. In this way, $\rho=\rho_m+\rho_f$ and $p=p_m+p_f$ again become the effective energy density and pressure in our model.

The field equations in $f(R, T)=R+2f(T)$ gravity with the system of units  $8\pi G=1=c$, are obtained as
\begin{equation}
R_{\mu\nu}-\frac{1}{2}R g_{\mu\nu}=T_{\mu\nu}+2 (T_{\mu\nu}+p g_{\mu\nu})f'(T)+f(T) g_{\mu\nu}.
\end{equation}
Adhav (2012) considered the simplest case $f(T)=\lambda T$, i.e., $f(R,T)=R+2\lambda T$, where $T=g^{\mu\nu}T_{\mu\nu}=\rho_m-3p_m$ for which the field equations (15) reduce to
\begin{equation}
R_{\mu\nu}-\frac{1}{2}R g_{\mu\nu}=(1+2\lambda) T_{\mu\nu}+\lambda(\rho_m-p_m) g_{\mu\nu}.
\end{equation}
The above field equations for the metric (1), yield
\begin{eqnarray}
  \left(\frac{\dot B}{B}\right)^2+2\frac{\dot A\dot B}{A B}&=&(1+3\lambda)\rho_m-\lambda p_m,\\
2\frac{\ddot B}{B}+\left(\frac{\dot B}{B}\right)^2&=&-(1+3\lambda)p_m+\lambda \rho_m,\\
\frac{\ddot A}{A}+\frac{\ddot B}{B}+\frac{\dot A \dot B}{AB}&=&-(1+3\lambda)p_m+\lambda \rho_m.
\end{eqnarray}
It is to be noted that the last term of equation ``(2.5)" in Ahdav's work  is ``$+2p\lambda$" which should be ``$-\lambda p$".

Using (13), (14) in (17) and (18), for $\lambda=-1/2$ and $\lambda=-1/4$, we obtain
\begin{eqnarray}
  \rho_m&=&\frac{3 k^2}{1+4 \lambda }-\frac{\beta ^2 e^{-6 k t}}{3(1+2 \lambda) },\\
  p_m&=&-\frac{3 k^2}{1+4 \lambda }-\frac{\beta ^2 e^{-6 k t}}{3(1+2 \lambda)}.
\end{eqnarray}
These are the correct expressions for the energy density and pressure which are different from those obtained by \cite{AdhavASS2012}. In both of these expressions the variable term decreases with time, consequently, the energy density and pressure increase with the cosmic evolution and attain a constant value $\rho_m=3 k^2/(4 \lambda +1)=-p_m$ as $t\to\infty$, while both physical quantities are infinite in the  infinite past.

The energy density for any physically viable cosmological model must be positive. It is clear from (20) that $\rho_m$ can be positive always if $1+4\lambda>0$ and $1+2\lambda<0$, but it is not possible. Similarly, the models with $-1/2<\lambda<-1/4$ also become physically unrealistic as $\rho_m$ remains negative always in this case. However, the energy density can be positive for some restricted times under the constraints
\begin{equation}
  t\leq\frac{1}{k}\ln\left[\frac{(1+4\lambda) \beta^2}{9(1+2\lambda) k^2}\right]^\frac{1}{6}\;\; \text{if}\;\; \lambda<-\frac{1}{2},
\end{equation}
and
\begin{equation}
  t\geq\frac{1}{k}\ln\left[\frac{(1+4\lambda) \beta^2}{9(1+2\lambda) k^2}\right]^\frac{1}{6} \;\; \text{if}\;\; \lambda>-\frac{1}{4}.
\end{equation}
The assumption (9)--(11) corresponds to an accelerated universe ($q=-1$). However, the acceleration may be an early inflation or a late time acceleration. Since the model with $\lambda<-1/2$ is physically viable only during early evolution, the acceleration must be an early inflation, whereas the model with $\lambda>-1/4$ is physically viable only at late times, the acceleration must be the present accelerating phase.

The main objective now remains to identify the nature of matter. The EoS parameter which is defined as $\omega_m=p_m/\rho_\mathfrak{m}$, gives
\begin{equation}
  \omega_m=-1+\frac{2 }{1+\gamma e^{6 \beta t}},
\end{equation}
where $\gamma=-9(1+2\lambda)k^2/(1+4\lambda)\beta^2$. The expression (24) looks similar to the EoS parameter of the effective matter in GR \cite{SinghBeeshamGRG2019}. The only difference is that here $\gamma$ contains a term $\lambda$ of $f(R,T)=R+2\lambda T$ gravity. The constraints obtained in (22) and (23) ensure the positivity of $\gamma$. Since $\omega_m\to1$ as $t\to-\infty$, and $\omega_m\to-1$ as $t\to\infty$, the matter behaves as stiff matter in the infinite past while it plays the role of a cosmological constant at late times. We see that $\omega_m$ diverges at a time $t=\frac{1}{k}\ln\left[\frac{(1+4\lambda) \beta^2}{9(1+2\lambda) k^2}\right]^\frac{1}{6}$, so it cannot be used to depict the behavior of the matter during intermediate evolution.

The early and late behavior of primary matter in our model matches with the characteristics of a scalar field. Due to the domination of kinetic energy over the scalar potential at early times, the scalar field acts like stiff matter. A scalar field with a self-interacting potential, due to the domination of the potential term over the kinetic term, gives rise to a negative pressure for driving super fast expansion during inflation. When the field enters into a regime in which the potential energy once again takes over from the kinetic energy, it exerts the same stress as a cosmological constant at late times, which happens however with a different energy density (in comparison to inflation). Therefore, let us substitute the primary matter with a scalar field (quintessence or phantom) for the further investigation.

\subsection{Scalar field model}
\label{sec:5}
The energy density and pressure of a minimally coupled normal ($\epsilon=1$) or phantom ($\epsilon=-1$) scalar field, $\phi$ with self-interacting potential, $V(\phi)$ are, respectively, given by
\begin{eqnarray}
   \rho_\phi&=&\frac{1}{2}\epsilon\dot\phi^2+V(\phi),\\
   p_\phi&=&\frac{1}{2}\epsilon\dot\phi^2-V(\phi).
\end{eqnarray}
Replacing $\rho_m$ with $\rho_\phi$ and $p_m$ with $p_\phi$, and using (25) and (26) in (20) and (21), the kinetic energy and scalar potential, respectively, are obtained as
\begin{eqnarray}
  \frac{1}{2}\epsilon\dot\phi^2&=&-\frac{\beta^2 e^{-6 k  t}}{3(1+2\lambda)},\\
V(t)&=&\frac{3k^2}{1+4\lambda}.
\end{eqnarray}
From (27), for reality of the solutions we must have $\lambda<-1/2$ if $\epsilon=1$ and $\lambda>-1/2$ if $\epsilon=-1$. It is to be note that the requirement of positive kinetic energy is equivalent to obeying the null energy condition (NEC)\footnote{$\rho+p\geq0$}. Since we have already ensured the positivity of energy density under the constraints (22) and (23), the WEC is satisfied for $\lambda<-1/2$ and $\lambda>-1/2$. In addition, the scalar potential also must be positive for a physically viable model, which is possible only for $\lambda>-1/4$. Hence, the model is consistent with a phantom scalar field only. It is to be noted that having a negative scalar potential is equivalent to violating the dominant energy condition (DEC)\footnote{$|\rho|\geq p$ or $\rho+p\geq0$ and $\rho-p\geq0$}. Furthermore,
the energy density for $\lambda>-1/4$, is positive only after a time (23), therefore, the model accommodates a late time acceleration only and it excludes the possibility of early inflation.

On integration (27), we get
\begin{equation}
  \phi=\phi_0\pm\left[\frac{2}{3(1+2\lambda)}\right]^\frac{1}{2}\frac{\beta e^{-3 k  t}}{3k},
\end{equation}
where $\phi_0$ is a constant of integration. Only the positive sign is compatible for physical consistency, so we proceed further with positive sign only.

The energy density and pressure of the phantom scalar field are given by (20) and (21), which can be expressed in terms of red shift, $z$ via the relation $a=a_0/(1+z)$  as
\begin{eqnarray}
  \rho_\phi&=&=\frac{3 k^2}{4 \lambda +1}-\frac{\beta ^2 (1+z)^{6}}{6 \lambda +3},\\
  p_\phi&=&-\frac{3 k^2}{4 \lambda +1}-\frac{\beta ^2 (1+z)^{6}}{6 \lambda +3},
\end{eqnarray}
respectively, where we have assumed the present scale factor to be unity, i.e., $a_0=1$.

Similarly, the constraints (22) and (23), respectively, can be expressed as
\begin{eqnarray}
   z&\geq&\gamma^\frac{1}{6}-1\;\; \text{for}\;\; \lambda<-\frac{1}{2},\\
   z&\leq&\gamma^\frac{1}{6}-1\;\; \text{for}\;\; \lambda>-\frac{1}{4}.
\end{eqnarray}
The EoS parameter (24), takes the form
\begin{equation}
  \omega_\phi=\left[-1+\frac{2 }{1+\gamma (1+z)^{-6}}\right]^{-1}.
\end{equation}
As aforesaid, though $\gamma$ contains the parameter $\lambda$ of $f(R,T)=R+2\lambda$ gravity, but being a single parameter expression the above EoS is identical to the EoS of the effective matter in GR \cite{SinghBeeshamGRG2019}. Hence, the primary matter in this model behaves similar to the effective matter in GR as shown in Fig. 1. An interesting fact here is that though Fig. 1 is identical to one in Ref. \cite{SinghBeeshamGRG2019} but it represents a different matter content. In present context it describes a part of matter while in the model of GR it describes the effective matter. This difference can also be seen in the expression of scalar field (29) which involve the parameter $\lambda$ of $f(R,T)=R+2\lambda T$ gravity.
So we can examine how $f(R,T)$ gravity affect the evolution of scalar field by analyzing its variation against different values of $\lambda$. We pursue further with the approach that we have adopted in our recent study (\cite{SinghBeeshamGRG2019}).

The present value of the EoS parameter is
\begin{equation}
  \omega_\phi(z=0)=\frac{1+\gamma}{1-\gamma}.
\end{equation}
Combined results from cosmic microwave background (CMB) experiments with large scale structure (LSS) data, the $H(z)$ measurement from the Hubble Space Telescope (HST) and luminosity measurements of Type Ia Supernovae (SNe Ia), put the following constraints on the EoS: $-2.68 <\omega_0< -0.78$ \cite{MelchiorrietalPRD2003}. These bounds become more tight, i.e., $-1.45 <\omega_0< -0.74$ \cite{HannestadMorstellPRD2002}, when the Wilkinson Microwave Anisotropy Probe (WMAP) data is included (also see \cite{Alametal0311364}; \cite{AlcanizPRD0401231}). For these observational limits from (35), we get $\gamma>2.19$ for the former bounds, and $\gamma>5.44$ for the latter. Now we can depict the profile of the EoS parameter for some values of $\gamma$ consistent with these observational outcomes.

\begin{figure}[h]
\includegraphics[width=8cm]{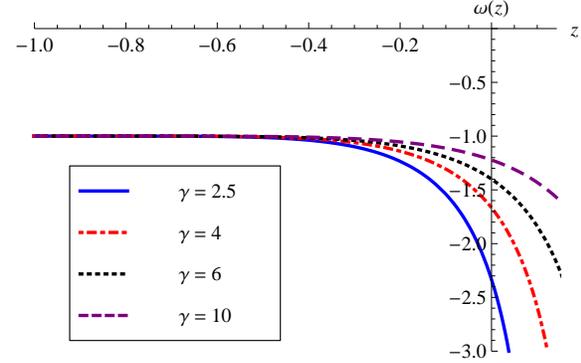}
  \caption{$\omega_\phi$ {\it \emph{\emph{versus}}} $z$ for different values of $\gamma$.}
  \label{fig:1}
\end{figure}

\noindent Figure 1 plots the behavior of the EoS parameter against redshift, under the constraint (33), for some values of $\gamma>2.19$. We see that $\omega_\phi<-1$, which confirms the theoretical outcome that the matter in the present model is of phantom type and becomes the cosmological constant in future.

The phantom scalar field (29) can be given as
\begin{equation}
  \phi=\phi_0\pm\left[\frac{2}{3\gamma(1+4\lambda)}\right]^\frac{1}{2}(1+z)^{3}.
\end{equation}

\begin{figure}[h]
  \includegraphics[width=8cm]{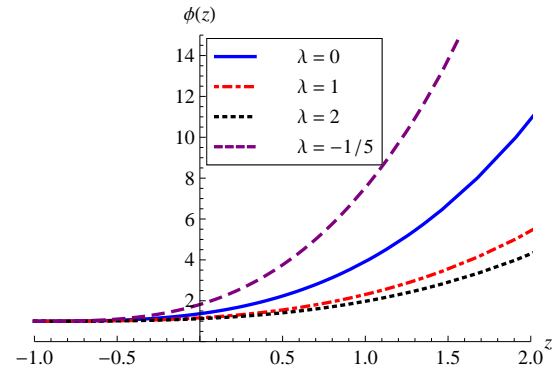}
\caption{$\phi(z)$ {\it \emph{\emph{versus}}} $z$ for different values of $\lambda$ with $\gamma=5$ and $\phi_0=1$.}
\label{fig:2}       
\end{figure}

\noindent The evolution of the scalar field against $z$ for some values of $\lambda$ with $\gamma=5$ (this value is consistent with the observational bounds mentioned above) and $\phi_0=0$ is shown in Fig. 2. The scalar field decreases from an infinite value with the evolution of the universe, and attains a finite minimum value at late times. If $\phi_0=0$, the scalar field vanishes at late times. The flat potential (28) can be identified as a cosmological constant. Moreover, if $\beta=0$ then $\phi=\phi_0$ and $\rho_\phi=3k^2=-p_\phi$, which essentially corresponds to a cosmological constant. If $\lambda=0$, the solutions reduces to the model one in GR \cite{SinghBeeshamGRG2019}. One may also readily verify that the standard de Sitter solutions for a flat Friedmann-Lemaitre-Robertson-Walker (FLRW) model of GR are recovered when $\beta=0$ and $\lambda=0$.

\subsection{The behavior of coupled matter}

Separating the energy densities and pressures of primary matter and coupled matter the field equations (17)--(19) can be written as
\begin{eqnarray}
  \left(\frac{\dot B}{B}\right)^2+2\frac{\dot A\dot B}{A B}&=&\rho_\mathfrak{m}+\rho_f,\\
2\frac{\ddot B}{ B}+\left(\frac{\dot B}{B}\right)^2&=&-(p_\mathfrak{m}+p_f),\\
\frac{\ddot A}{A}+\frac{\ddot B}{B}+\frac{\dot A \dot B}{AB}&=&-(p_\mathfrak{m}+p_f),
\end{eqnarray}
where $\rho_f=\lambda(3\rho_\mathfrak{m}-p_\mathfrak{m})$ and $p_f=\lambda(3p_\mathfrak{m}-\rho_\mathfrak{m})$, are the energy density and pressure of coupled matter, which are obtained as
\begin{eqnarray}
  \rho_f&=&\frac{12  \lambda k^2  }{4 \lambda +1}-\frac{2 \lambda \beta ^2   e^{-6 k t}}{6 \lambda +3},\\
  p_f&=&-\frac{12 \lambda  k^2}{4 \lambda +1}-\frac{2 \lambda \beta ^2   e^{-6 k t}}{6 \lambda +3}.
\end{eqnarray}
Although $\rho_f$ remains always positive for $-1/2<\lambda <-1/4$, but the energy density of primary matter becomes negative for these values, we exclude this case. For $\lambda<-1/2$, $\rho_f$ becomes negative at early times, we exclude this case too. Similarly, when $-1/4<\lambda<0$, it becomes negative at late times. Notwithstanding, $\rho_f$ for $\lambda>0$ is positive at late times. Hence, the model in this case provides a realistic scenario. The EoS parameter, $\omega_f=p_f/\rho_f$ diverges at $t=t_\star$, so it is not worthwhile using it to depict the behavior of coupled matter. Therefore, we shall study the nature of coupled matter through the energy conditions. We require
\begin{eqnarray}
  \rho_f+p_f&=&-\frac{4  \lambda\beta ^2  e^{-6 k t}}{6 \lambda +3},\\
  \rho_f-p_f&=&\frac{24 \lambda k^2  }{4 \lambda +1}.
\end{eqnarray}
The NEC and DEC can be satisfied, respectively, for
\begin{eqnarray}
  -\frac{1}{2}&<&\lambda <0, \\
 \lambda &<&-\frac{1}{4} ,\;\; \text{or}\;\; \lambda>0.
\end{eqnarray}
Since $\rho_f$ is positive only for $\lambda>0$, therefore, the coupled matter violates the NEC but holds the DEC. Hence, the behavior of coupled matter is similar to the primary matter, i.e., it also behaves as phantom DE. Thus, the behavior of coupled matter also shows that the model is viable to describe a late time cosmic acceleration in presence of phantom matter.

\section{Conclusion}

\cite{AdhavASS2012} studied an LRS Bianchi I model with constant expansion rate in $f(R,T)=R+2\lambda T$ gravity. The solutions obtained by the author are mathematically and physically invalid due to an incorrect field equation. We have reconsidered this model in the present paper. We have also extended the solutions to a scalar field (quintessence or phantom) model. While \cite{AdhavASS2012} discussed only the kinematical behavior of the model, we have also explored the physical properties in detail keeping the physical viability of the solutions at the center. Notwithstanding, a wrong field equation, the behavior of the geometrical parameters is not affected. Moreover, the kinematical parameters in such formulation do not depend on $f(R,T)$ gravity. Firstly, we have shown that the geometrical behavior remains the same as in GR. We have pointed out that \cite{AdhavASS2012} misunderstood the time of origin of the universe. The author discussed the evolution from $t=0$ to $t\to\infty$. The author probably understood the origin of the universe at $t=0$, while the model has an infinite past. For details, the readers may refer our recent works  \cite{SinghBeeshamGRG2019,SinghBeeshamEPJP2020}. The solutions in $f(R,T)$ gravity model are valid for all values of $\lambda$ except $\lambda\neq-1/2$ and $\lambda\neq-1/4$. The assumption of constant expansion rate gives a constant value of the deceleration parameter, $q=-1$. Consequently, the model can describe only an accelerating universe. However, the acceleration may be an early inflation or present acceleration. To ensure this we have obtained the constraints for a physically realistic scenario and found that the model is viable to describe  only the present accelerating phase.

The evolution of the universe is governed by the effective matter. The behavior of the effective matter has already been studied by us elsewhere recently \cite{SinghBeeshamGRG2019}. The $f(R,T)$ gravity does not affect the behavior of the effective matter and it remains the same as one in GR. It is important to mention here that $f(R,T)$ gravity does not alter the behavior of effective matter in the formulations where the kinematical behaviour is fixed by some geometrical parameters. The effective matter in the present study thus also remains the same as in GR \cite{SinghBeeshamGRG2019}. In present study, an extra matter (different from the primary matter) appears due to the coupling between matter and $f(R,T)$ gravity. We have explored the characteristics of primary matter as well as of coupling matter.

The primary matter in this model acts similar to the effective matter in GR. When primary matter is replaced with a scalar field (normal or phantom) model, the model has been found consistent only with a phantom scalar field. The phantom field decreases with the cosmic evolution, while the scalar potential remains flat throughout the comic evolution. The scalar potential can be thought as cosmological constant. The model can also be consistent with normal scalar field, but the scalar potential becomes negative in that case which would be unrealistic. The coupled matter behaves similar to primary matter. The viable models are possible only for $\lambda>0$.

We have also examined the consistency of the behavior of primary matter with the the observational data by borrowing some current values of the EoS parameter from some observational outcomes. The dynamics of the EoS parameter supports the observational results and suggests that the phantom field has started dominating over the other energy contents somewhere between $0.2\lesssim z\lesssim0.5$. The scalar field model also evidences that if one demands an accelerating cosmic expansion from an anisotropic model, then the model represents a viable cosmological scenario (obeying NEC and WEC) only after a time when the universe enters into an accelerating phase.

It is to be noted that \cite{ShamirJETP2014} obtained solutions of the general Binachi I model with  constant expansion rate in $f(R,T)=R+2\lambda$ gravity. Those solutions are also valid for late times only as the energy density is negative at early times. Hence, our results can also be interpreted within the general Binachi I spacetime model. We believe that only the kinematical behavior would be different, but the physical behavior will remain the same.

As the model is capable to explain the present cosmic acceleration without the use any hypothetical exotic matter, $f(R,T)=R+2\lambda T$ gravity can be a good alternative to GR.

\bibliographystyle{spr-mp-nameyear-cnd}
\makeatletter
\let\clear@thebibliography@page=\relax
\makeatother

\end{document}